\documentclass[%
 reprint,
 amsmath,
 amssymb,
 aps,
 floatfix,
]{revtex4-2}

\usepackage{graphicx}% Include figure files
\usepackage{dcolumn}% Align table columns on decimal point
\usepackage{bm}% bold math
\usepackage[hidelinks]{hyperref}
\usepackage[dvipsnames]{xcolor} %Allows coloured text
\usepackage{xr}%cross referencing with hyper links

\newcommand*{\changes}[1]{\textcolor{black}{#1}} %of text within 'changes' command

\makeatletter

\begin{document}

\preprint{APS/123-QED}

\title{Spin polarised quantised transport via one-dimensional nanowire-graphene contacts}% Force line breaks with \\

\author{Daniel Burrow$^{1}$, Jesus C. Toscano-Figueroa$^{1,2}$, Victor H. Guarochico-Moreira$^{1,3}$, Khalid Omari$^{1,4}$, Irina V. Grigorieva$^{1}$, Thomas Thomson$^{5}$, and Ivan J. Vera-Marun$^{1}$}

\email{Correspondence to ivan.veramarun@manchester.ac.uk}

\affiliation{$^{1}$Department of Physics and Astronomy, University of Manchester, Manchester M13 9PL, United Kingdom} 

\affiliation{$^{2}$Consejo Nacional de Ciencia y Tecnología (CONACyT), Av. Insurgentes Sur 1582, Col. Crédito Constructor, Alcaldía Benito Juarez, C.P. 03940, Ciudad de México, México} 

\affiliation{$^{3}$Facultad de Ciencias Naturales y Matemáticas, Centro de Investigación y Desarrollo en Nanotecnología, Escuela Superior Politécnica del Litoral, ESPOL, Campus Gustavo Galindo, Km. 30.5 Vía Perimetral, Guayaquil, 090902, Ecuador} 

\affiliation{$^{4}$ The Open University, Walton Hall, Kents Hill, Milton Keynes, MK7 6AA}

\affiliation{$^{5}$Nano-Engineering and Spintronic Technologies (NEST), Department of Computer Science, University of Manchester, Manchester M13 9PL, United Kingdom}

\date{\today}

\begin{abstract}
Graphene spintronics offers a promising route to achieve low power 2D electronics for next generation classical and quantum computation. As device length scales are reduced to the limit of the electron mean free path, the transport mechanism crosses over to the ballistic regime. However, ballistic transport has yet to be shown in a graphene spintronic device, a necessary step towards realising ballistic spintronics. Here, we report ballistic injection of spin polarised carriers via one-dimensional contacts between magnetic nanowires and a high mobility graphene channel. The nanowire-graphene interface defines an effective constriction that confines charge carriers over a length scale smaller than that of their mean free path. This is evidenced by the observation of quantised conductance through the contacts with no applied magnetic field and a transition into the quantum Hall regime with increasing field strength. These effects occur in the absence of any constriction in the graphene itself and occur across several devices with transmission probability in the range $T = 0.08 - 0.30$.
\end{abstract}

\maketitle

\section{Introduction} \label{sec:Intro}

Spin based electronics (spintronics) offers a promising route to achieving low power and high speed computation beyond CMOS technology \cite{Wolf-2001, Awschalom-2002, Zutic-2004, Behin-Aein-2010, Hu-2018, Dieny-2020, Li-2022}, and potentially to energy efficient spin qubits for applications in quantum information processing \cite{Leuenberger-2001, Tong-2021, Niknam-2022}. Graphene, with its high carrier mobility and gate tuneable charge carrier density \cite{Novoselov-2004, CastroNeto-2009}, is an ideal choice of material for studying spin transport \cite{Avsar-2020, Ahn-2020}. There have been numerous experimental demonstrations of graphene-based quantum devices built upon various architectures, \changes{including quantum dots \cite{Recher-2009, Eich-2018, Tong-2021, Banszerus-2022} and quantum point contacts \cite{Terres-2016, Overweg-2018, Overweg-2018b, Lee-2020, Kun-2020}}, which can enable further development of quantum technology. Yet, such architectures have not been employed within graphene spintronic devices. 
%where they could be exploited to improve important aspects such as efficient generation and manipulation of spin signals \cite{Behin-Aein-2010, Hu-2018, Dieny-2020, Li-2022}.

Modern graphene spintronic devices typically use hexagonal boron nitride (hBN) to support and encapsulate the graphene channel, providing it with an atomically flat substrate, enhancing its charge and spin transport capabilities \cite{Deshpande-2009, Dean-2010, Xue-2011, Zomer-2011, Mayorov-2011, Guimaraes-2014, InglaAynes-2015, InglaAynes-2016, Laturia-2018, Gurram-2018} relative to earlier devices on Si/SiO$_{2}$ substrates \cite{Tombros-2007, Han-2011, Vera-Marun-2011}. As the quality of graphene devices is improved, the electronic mean free path can reach up to $\mu$m scales, approaching or exceeding device dimensions and giving rise to ballistic transport \cite{Tombros-2011, Guimaraes-2012, Terres-2016, Banszerus-2016, Overweg-2018, Kun-2020}. This is of particular interest in the field of low power spintronics due to suppressed scattering in the ballistic regime, and the potential for coherent control over spin transport \cite{Datta-1990, Vila-2020}.  However, at present there are very few experimental studies regarding ballistic spintronics in graphene, despite this being the focus of studies in low dimensional nanostructures based on III-V semiconductors \cite{Grundler-2001, Kravchenko-2003, Chen-2015, Ciorga-2016}, carbon nanotubes \cite{Sahoo-2005, Man-2006, Crisan-2016}, and evidence of spin polarised edge states in graphene nanoribbons \cite{Prudkovskiy-2022}.

\begin{figure*}[htb]
    \centering
    \includegraphics[width = \textwidth]{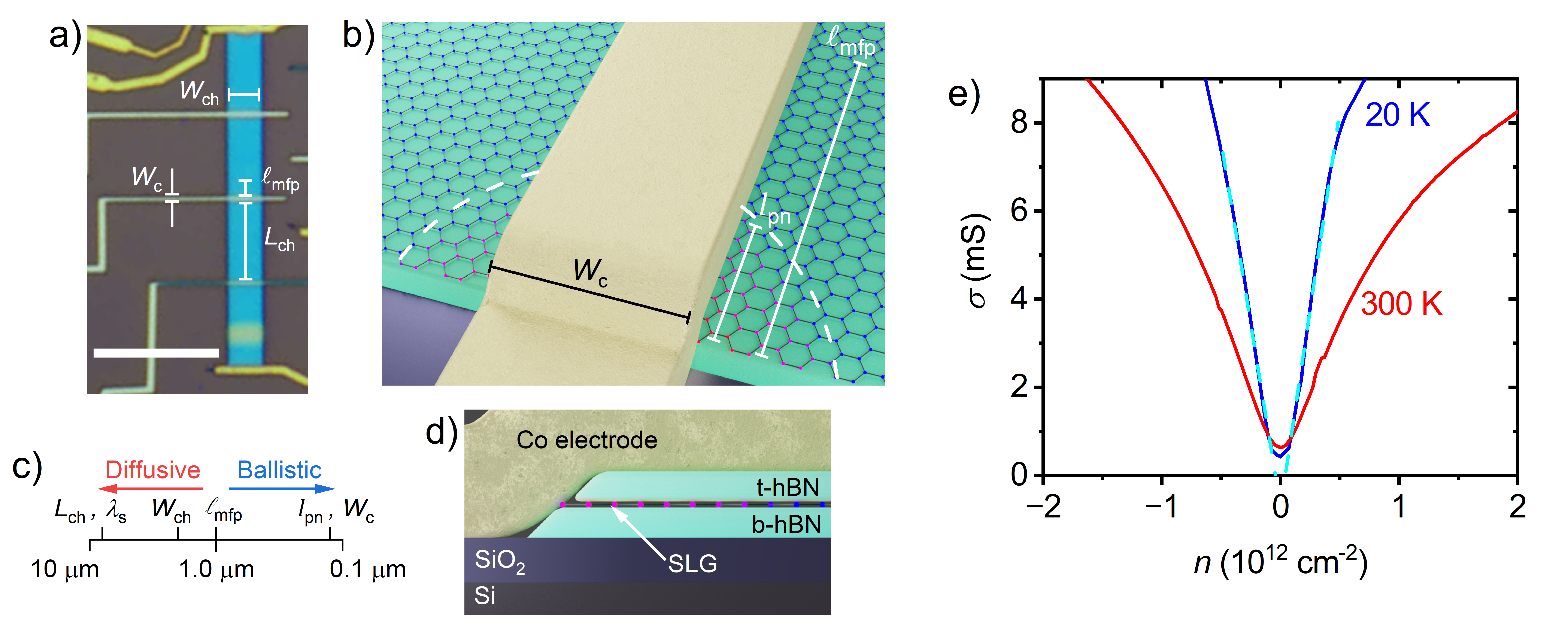}
    \caption{\small Device and 1D contact architecture. \textbf{(a)} Optical microscope image of a representative graphene spintronic device (D3) on SiO$_2$ substrate. Vertical blue strip is the hBN-encapsulated graphene channel, pale yellow bars are ferromagnetic cobalt (Co) electrodes and bright yellow bars are gold (Cr/Au) reference contacts. $W_{\text{c}}$ is contact width, $W_{\text{ch}}$ is channel width, and $L_{\text{ch}}$ is channel length (i.e. distance between adjacent contacts). Note $W_{\text{c}}$ is varied throughout the device such that each contact has a different magnetic coercivity. Scale bar (bottom left) is 10 $\mu$m. \textbf{(b)} Illustration of the nanowire-graphene interface. Red colouring within the dashed white line represents $n-$doping relative to the rest of the channel, due to charge transfer from the Co nanowire. The doping length scale, $l_{\text{pn}}$, is less than the mean free path of injected charge carriers, $\ell_{\text{mfp}}$. Edge termination in this illustration does not reflect that of real devices, which have disordered edges. \textbf{(c)} Breakdown of device length scales indicating where transport can be considered ballistic. $\lambda_{\text{s}}$ is the spin diffusion length. \textbf{(d)} Side profile view of the bottom hBN (b-hBN)/single layer graphene (SLG)/top hBN (t-hBN) stack. \textbf{(e)} Conductivity of device D1 vs carrier density in the channel at room temperature (300 K) and low temperature (20 K). Dashed cyan lines are linear fits to the conductivity, used to estimate the field-effect mobility of the channel.}
    \label{fig:Schematic}
\end{figure*}

In the case of an encapsulated channel, metal electrodes can be placed over the hBN/graphene/hBN stack to form a one-dimensional (1D) contact \cite{Wang-2013a, Xu-2018b, Choi-2022} where the metal bonds with the exposed graphene edge. 1D contacts have been shown to be far less invasive than direct, and tunnel barrier, top contacts, where a large area of the graphene surface bonds to the electrode \cite{Rashba-2000, Huard-2008, Mueller-2009, Maassen-2012, Allain-2015, Karpiak-2017, Xu-2018b, Guarochico-Moreira-2022}. We recently demonstrated fully encapsulated graphene channels with nanoscale magnetic 1D contacts that combine high charge mobility and mean free paths in excess of $\sim 1\, \mu$m, with efficient spin injection and spin diffusion lengths up to $\sim 20\, \mu$m \cite{Guarochico-Moreira-2022}. The separation between contacts in these devices was on the order of $\geq 10\, \mu$m, while the channel widths were typically $1-2\, \mu$m, implying spin transport through the channel occurs in the diffusive regime \cite{Zutic-2004}. 

In this work, we present observations on the injection of spin polarised carriers via 1D magnetic nanowire-graphene interfaces with widths on the order of $\sim 100$ nm (Figure \ref{fig:Schematic}a-d). We find that carriers are confined on a length scale below their mean free path upon injection to the channel, causing the emergence of quantised energy subbands in their band structure, evidenced by quantised conductance through the 1D contacts in the absence of a magnetic field. Contact conductance is analysed with the Landauer equation to quantitatively evaluate the effective constriction width, defined by the magnetic nanowire, and the energy scale of the observed subbands. Both extracted values agree with ballistic injection with transmission probability in the range $T=0.08-0.30$, orders of magnitude higher than tunnel barrier contacts and comparable to values found in studies regarding edge contacted graphene \cite{Matsuda-2010, Nath-2016, Ben-Shalom-2016}. Further evidence of the ballistic nature of transport via the 1D magnetic contacts is explored by probing the transition into the quantum Hall regime, with magnetic fields as low as $|B|_{\perp} \leq 1$ T, yielding a consistent analysis of the 1D interface characteristics. Crucially, we obtain these results without the need to engineer any physical constriction within the graphene itself,overcoming nanofabrication challenges associated with defining quantum point contacts \cite{Tombros-2011, Terres-2016, Overweg-2018, Kun-2020}. The demonstration of quantised transport via spin polarised injectors to a graphene channel represents the first step towards the realisation of ballistic graphene spintronic devices. 

\section{Results and Discussion}
\label{sec:R&D}

\subsection{Charge transport} \label{sec:Charge}

Results presented in this work are taken from four devices: D1 - D4. All devices comprise fully encapsulated graphene channels employing nanoscale 1D contacts (a summary of the fabrication process and measurement set up can be found in Methods, while the physical properties of each device can be found in Supplementary Table SI). All data are recorded at a temperature of 20 K, unless stated otherwise. Figure \ref{fig:Schematic}e shows the conductivity, $\sigma$, against $n$, where the neutrality (Dirac) point, $n=0$, is defined by the back gate voltage at which $\sigma$ is a minimum. Applying a linear fit to this data at low carrier density, we extract the field effect mobility, $\mu = 1/e(\text{d}\sigma/\text{d}n)$. Around $n \sim 2 \times10^{11}$ cm$^{-2}$, the extracted values are $\mu = 43,000 \pm 2,000$ cm$^2$V$^{-1}$s$^{-1}$ at 300~K and $\mu = 105,000 \pm 5,000$ cm$^2$V$^{-1}$s$^{-1}$ at 20~K. These values are significantly higher than those extracted from graphene spintronic devices of other architectures \cite{Han-2011, Guimaraes-2014, Xu-2018b}, but consistent with the highest mobilities found in devices employing full encapsulation and 1D contacts \cite{Guarochico-Moreira-2022}. Lower mobility at room temperature (300~K) indicates transport is dominated by electron-phonon scattering, with contributions from intrinsic graphene phonons and the surface optical mode of the hBN substrate \cite{Konar-2010, Li-2015}, while at low temperature (20~K) the narrow Dirac peak evidences a low impurity density in the channel of $<5\times10^{10}$ cm$^{-2}$ \cite{Dean-2010}.
Conductivity is related to the mean free path of carriers, $\ell_{\text{mfp}}$, via the Einstein relation, $D=\sigma/e N$, where $D = v_{\text{F}} \ell_{\text{mfp}}/2 $ is the diffusion coefficient, $N$ is the density of states of graphene and $v_{\text{F}}$ is the Fermi velocity (see Methods). At $T=20$ K, we extract a value of $\ell_{\text{mfp}} = 0.74 \pm 0.07\, \mu$m at a carrier density $|n| = 5 \times10^{11}$ cm$^{-2}$ (Supplementary Figure S2c). The value of $\ell_{\text{mfp}}$ is far below the separation between contacts, $L_{\text{ch}}=12.2$ $\mu$m, indicating that charge transport measured along the channel is diffusive.

The resistance of the 1D contacts as a function of $V_{\text{BG}}$ is measured by employing the 3-terminal configuration (Figure \ref{fig:Ballistic}c). All contacts display an electron-hole asymmetry, with higher transmission in the electron regime. We also observe a small shift between the contact conductance minimum, used to define the contact carrier density, $n_{\text{c}}$, and the graphene channel neutrality point, with the contact conductance minimum shifted towards negative carrier density by $\sim 1\times10^{11}$cm$^{-2}$ (Supplementary Figure S2d). This implies $n-$type doping of graphene adjacent to the 1D interface, arising due to charge transfer from the cobalt nanowire. The $n-$doped region at the nanowire-graphene interface forms a tuneable potential profile that extends into the channel over a length scale, $l_{\text{pn}}\leq 100$ nm (Figure \ref{fig:Schematic}b) \cite{Giovannetti-2008, Huard-2008, Mueller-2009, Xia-2011, Khomyakov-2009, Asshoff-2017}. The extracted value of $\ell_{\text{mfp}}$ is $2-3\times$ below the physical width of the channel, $W_{\text{ch}}=2\, \mu$m, but is much greater than $l_{\text{pn}}$. Considering this, contact resistance consists of dominant contributions from scattering at the rough nanowire-graphene interface, and from the surrounding potential profile. 

The contacts display plateaus in their resistance as a function of $V_{\text{BG}}$ (Figure \ref{fig:Ballistic}c and Supplementary Figure S2d). As the value of $\ell_{\text{mfp}}$ is greater than the widths the 1D contacts ($W_{\text{c}} = 150-350$~nm), the observed plateaus in $R_{\text{c}}$ are explained by the formation of subbands in the energy spectrum of injected charge carriers, due to confinement through the 1D nanowire-graphene interface. The combination of confinement from the 1D constriction defined by the magnetic nanowire-graphene interface, and the surrounding tuneable potential profile, forms an effective tuneable quantum point contact through which transport occurs ballistically. From this point on, we will refer to these as edge quantum point contacts (e-QPCs).

\subsection{Zero field ballistic transport through nanoscale 1D contacts}\label{sec:Spectro} 

\begin{figure*}[htb!]
    \centering
    \includegraphics[width=\textwidth]{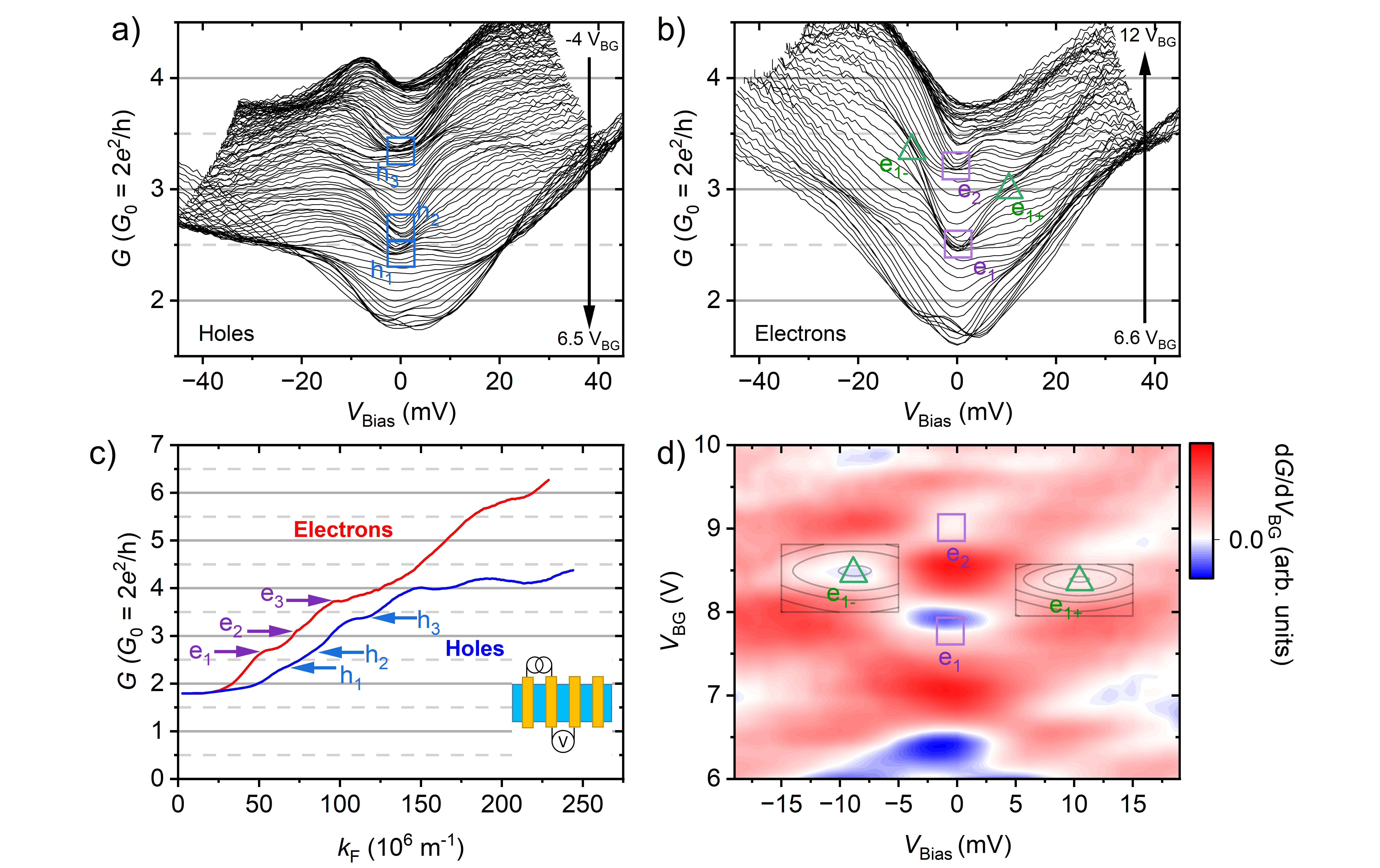}
    \caption{\small Zero field quantised conductance via e-QPCs (device D2). \textbf{(a)} Hole and \textbf{(b)} electron conductance as a function of bias voltage, $V_\text{Bias}$, applied between the magnetic electrode and the graphene channel. Each black line represents a bias sweep at a different fixed back gate voltage $V_{\text{BG}}$; the conductance minimum for this e-QPC occurs at $6.6 {V_{\text{BG}}}$. Blue squares in (a) highlight zero bias hole subbands ($h_{1,2,3})$. Likewise, purple squares in (b) show zero bias electron subbands, ($e_{1,2}$), while green triangles highlight finite bias plateaus ($e_{1-,1+}$). \textbf{(c)} Conductance vs $k_{\text{F}}$ ($k_{\text{F}} = \sqrt{\pi n_{\text{c}}})$. This data represents a separate measurement, but can be thought of as a trace of the conductance along $V_{\text{Bias}}=0$ in the spectroscopy data. Arrows indicate electron and hole subbands (corresponding to squares in panels (a) and (b)). Inset shows a schematic of the 3-terminal measurement configuration. \textbf{(d)} Spectroscopy data for electron doping (panel (b)), plotted as a 2D map of transconductance against $V_{\text{Bias}}$ and $V_{\text{BG}}$. Plateaus occur where $\text{dG}/\text{dV} = 0$, which is represented in white. Black boxes represent the area fit with a 2D Lorentzian function, to find the local minima and quantify the position of the finite bias plateaus ($e_{1+,1-}$).}
    \label{fig:Ballistic}
\end{figure*}

Figure \ref{fig:Ballistic}c presents contact conductance, $G$, extracted from an e-QPC, in units of $G_0=2e^2/h$ and plotted as a function of Fermi wave vector, $k_{\text{F}} = \sqrt{\pi n_{\text{c}}}$, where $n_{\text{c}}$ is the e-QPC carrier density.
The data shows a series of periodic plateau-like features, which emerge at low temperature (arrows). Such features are caused by the quantisation of available energy states in the e-QPC into discrete 1D subband \cite{Beenakker-1991, Datta-1995}. This is a fingerprint of ballistic conduction, indicating that carriers are confined by the 1D magnetic nanowire-graphene interface upon injection to the channel. Conductance in the ballistic regime is governed by the Landauer equation, which, for graphene, can be approximated as \cite{Terres-2016},
\begin{equation}
    G = \frac{4 e^2}{h}MT = \frac{4 e^2}{h}\frac{k_{\text{F}} W_{\text{c}}}{\pi}T,
    \label{eq:Landauer}
\end{equation} %\noindent
where $e^2/h$ is the conductance quantum, $T$ is the transmission probability, and $M$ is the mode number, defined as the number of half Fermi wavelengths, $\lambda_{\text{F}}/2 = \pi/k_{\text{F}}$, that fit within the constriction width, $W_{\text{c}}$ \cite{Beenakker-1991, Datta-1995}. The factor of 4 is unique to graphene and reflects its four-fold degeneracy, coming from the 2 spin and 2 valley states that its carriers can occupy. $T$ takes a value between 0 and 1, and modifies the magnitude of conductance to account for carriers that are back-scattered, for example off rough edges. The value of $T$ can be estimated from the spacing of consecutive 1D subbands, relative to the ideal case, when $T=1$ and plateaus are spaced by $4e^2/h$. Subbands emerging in Figure \ref{fig:Ballistic}c are separated by $\sim e^2/h$, a factor of $\sim4$ reduction from the ideal case. Values of $T$ extracted by this method, from all devices, are in the range $T=0.08-0.30$ (see Supplementary Table SI). 

\begin{figure*}[htb]
    \centering
    \includegraphics[width=\textwidth]{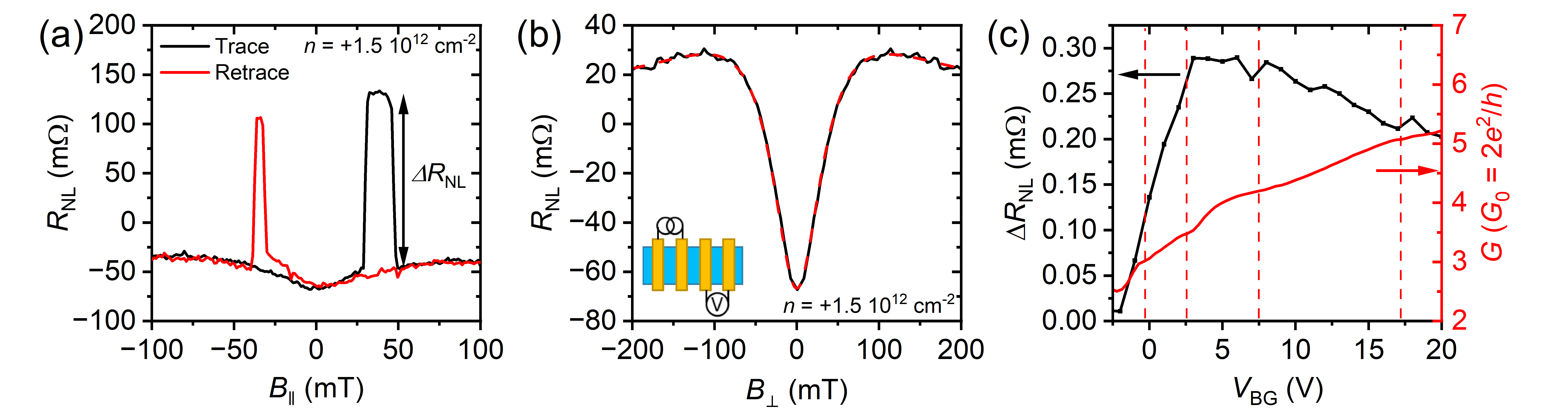}
    \caption{\small Spin transport, device D1. \textbf{(a)} Spin valve, at a channel density of $n=1.5\times 10^{12}$ cm$^{-2}$, taken by sweeping magnetic field along the easy axis of the contacts. Trace implies $B$ is swept from negative to positive, while retrace implies the opposite direction. The magnitude of the spin signal, $\Delta R_{\text{NL}}$, is indicated by the black arrow. \textbf{(b)} Hanle spin precession measurement, at a channel density of $n=1.5\times 10^{12}$ cm$^{-2}$, taken by sweeping a magnetic field out of plane. The line shape signifies spin moments have almost fully precessed within 200 mT. Dashed red line indicates a fit to the data using the solutions to the 1D Bloch equation. Inset shows schematic of non-local configuration used for spin transport measurements. \textbf{(c)} Evolution of $\Delta R_{\text{NL}}$ with increasing $V_{\text{BG}}$ (black line, left axis), superimposed with conductance through the injector (red line, right axis), for electron transport only. Dotted red lines indicate plateaus in the conductance that correspond to energy subbands.}
    \label{fig:Spin_transport}
\end{figure*}

As the transmission probability is crucial to the Landauer description of ballistic conduction, we verify our estimation of $T$ by performing finite bias spectroscopy on the e-QPCs, where conductance at a fixed carrier density is measured as a function of a bias voltage applied between the electrode and channel (see Methods) \cite{Kouwenhoven-1989}. Conducting bias sweeps at different back gate voltages allows us to generate bias maps for the hole (Figure \ref{fig:Ballistic}a) and electron (Figure \ref{fig:Ballistic}b) transport regimes, which help to visualise the energy subbands. These can be seen at points along zero bias where the lines are closely spaced, or overlapping, with their positions agreeing with the plateaus observed in Figure \ref{fig:Ballistic}c.

Transmission across the magnetic nanowire-graphene interface is determined by the coupling of graphene $p-$orbitals to the surface $d-$orbitals of the ferromagnetic electrodes. Hence, there is a mismatch in work functions on either side of the interface, and it is expected that $T<1$ \cite{Matsuda-2010, Xia-2011, Khoo-2016}. In fact, $T$ across the e-QPCs is somewhat tuneable via the back gate voltage, owing to the tuneable potential profile surrounding the 1D interface. Close to the interface, the graphene is $n$-type doped due to charge transfer from the magnetic nanowire. However, the Fermi energy of the graphene channel is free to change as the back gate voltage is varied \cite{Giovannetti-2008, Huard-2008, Mueller-2009, Khomyakov-2009, Asshoff-2017}. Hence, in the electron doping regime, the e-QPC has an $n$-$n$ potential profile, scattering is reduced and $T$ is increased. Meanwhile, in the hole doping regime, the e-QPC has an $n$-$p$ profile, which increases scattering and reduces $T$ in this region \cite{Xia-2011, Terres-2016}. Consequently, the plateaus in the hole doping regime appear smeared and occur at increasingly smaller spacings as hole density increases and $T$ decreases.  This behaviour is responsible for the observed divergence of conductance between electrons and holes for $k_{\text{F}} > 150\times10^6$ m$^{-1}$ (Figure \ref{fig:Ballistic}c). 

With the above considerations in mind, the following analysis is performed in the medium carrier density range $k_{\text{F}} \approx 50 - 150 \times10^6$ m$^{-1}$ or $|n_{\text{c}}| \approx 0.5 - 7.0\times 10^{11}$ cm$^{-2}$, where conductance is comparable between the two carrier regimes. Averaging the spacing between plateaus within this carrier density range, across the two data sets (Figures \ref{fig:Ballistic}a,b,c), yields a transmission of $T(e) = 0.26 \pm 0.04$ for electrons, and $T(h) = 0.23 \pm 0.05$ for holes. The asymmetry in these results is consistent with the interpretation of a $p$-$n$ junction in the adjacent graphene. With these estimations of $T$, equation \ref{eq:Landauer} can be used to evaluate the effective constriction width defined by the magnetic nanowire-graphene interface, which is  $W_{\text{c}}(e) = 240 \pm 20$ nm for electrons, and $W_{\text{c}}(h) = 220 \pm 20$ nm for holes. The width of the electrode used for this measurement is 300~nm. Hence, the extracted width for both carrier polarities agree within their uncertainty and fall within $20\%$ of the nominal physical width of the magnetic nanowire, providing confirmation that our estimate of the transmission probability is accurate. 

The effective constriction width can also be evaluated from the energy spacing between 1D subbands, which can be directly determined from spectroscopy data. Here, the non-linear conductance response is studied by identifying the bias value at which plateaus from two adjacent subbands merge \cite{Kouwenhoven-1989}. To more easily visualise this effect, Figure \ref{fig:Ballistic}d presents the bias spectroscopy data as a 2D colour map of the transconductance (d\textit{G}/d\textit{V\textsubscript{BG}}). When plotted in this way, a distinct diamond pattern is visible, formed from plateaus at zero bias (purple squares) and the finite bias point at which they merge (green triangles). Fitting a local minima to the crossing point (black boxes Figure \ref{fig:Ballistic}d) allows us to quantify the average energy spacing of the subbands, which gives $\Delta E = 9.5 \pm 1.4$ meV.

By combining the graphene dispersion relation, $E = \hbar v_{\text{F}} k_{\text{F}}$, where $v_{\text{F}} = 1\times10^{6}$ ms$^{-1}$ is the Fermi velocity of graphene, with the number of modes, $M = k_{\text{F}} W_{\text{c}}/\pi$, the energy of a given subband is given by,
\begin{equation}
    E = \frac{\hbar v_{\text{F}} \pi}{W_{\text{c}}} M.
    \label{eq:EnergySpace}
\end{equation} %\noindent
Employing Equation \ref{eq:EnergySpace}, along with the estimate of $\Delta E$, the effective constriction width is calculated as $W_{\text{c}}(e) = 220 \pm 30$ nm. This value agrees within the experimental error with that extracted by applying the Landauer equation to Figure \ref{fig:Ballistic}c. Thus, using two different methods, applied to separate data sets, we demonstrate a consistent quantitative picture of ballistic conduction through an e-QPC.

\subsection{Spin transport} \label{sec:Spin}

We demonstrate efficient spin injection via the e-QPCs, via non-local (inset Figure \ref{fig:Spin_transport}b) spin valve and spin precession (Hanle) signals, which are shown respectively in Figures \ref{fig:Spin_transport}a and \ref{fig:Spin_transport}b. The spin valve signal shows two anti-parallel states, one for negative $B$ field and one for positive, which occur due to an abrupt reversal in the magnetisation direction of a magnetic electrode, relative to its adjacent electrode \cite{Takahashi-2003, Zutic-2004, Tombros-2007, Popinciuc-2009}. These switches signify the injection and detection of spin currents via the e-QPCs, with a spin signal magnitude of $\Delta R_{\text{NL}} = 180 \pm 10$ m$\Omega$. Spin precession data shown in Figure \ref{fig:Spin_transport}b is fitted with the standard spin precession equation \cite{Zutic-2004, Tombros-2008, Swartz-2013, Ingla-Aynes-2021}, which yields a spin diffusion coefficient $D_{\text{s}} = 0.37 \pm 0.04$ m$^2$ s$^{-1}$, a spin lifetime $\tau_{\text{s}} = 170 \pm 14$ ps, a contact polarisation $P = 4.8 \pm 0.5 \%$, and a spin diffusion length $\lambda_{\text{s}} = 7.9 \pm 0.7\, \mu$m. These values compare favourably with those found across several of our previous devices \cite{Guarochico-Moreira-2022}, consistent with the high mobility found for this spintronic device.

We perform spin valve measurements over a gate voltage range where the clearest plateaus are observed in contact conductance ($-2.5$ -- $20$ V), extracting values of $\Delta R_{\text{NL}}$ at each $V_{\text{BG}}$ value (Figure \ref{fig:Spin_transport}c). The spin signal dependence is plotted alongside conductance, $G$, through the spin injecting contact, which displays several plateaus, arising from confinement, spaced by $\sim e^2/h$ in agreement with the previous measurement; plateaus are highlighted by dashed red lines, placed according to minima in d$G$/d$V_{\text{BG}}$ (Supplementary Figure S5b). Note that the first three of the plateaus seen in Figure \ref{fig:Spin_transport}c ($V_{\text{BG}}<10$ V), are also visible in a separate measurement of contact conductance taken from the same contact, during a subsequent cool down (Figure \ref{fig:FanAndSpectro}b). Hence, the observed low field confinement plateaus are reproducible across several thermal cycles. The first of these subbands occurs in a region of rapid increase in $\Delta R_{\text{NL}}$, explained in part by a similar decrease in mismatch parameter close to neutrality (Supplementary Figure S5a). Subsequent plateaus are located close to weak features in the spin signal, but more devices are needed to determine a connection between the presence of quantised transport and modulations of spin signal, as was found for carbon nanotube quantum dots with ferromagnetic contacts \cite{Sahoo-2005, Man-2006, Crisan-2016}.  

\subsection{Transition to quantum Hall regime} \label{sec:QHE}

\begin{figure}[tb]
    \centering
    \includegraphics[width=0.5\textwidth]{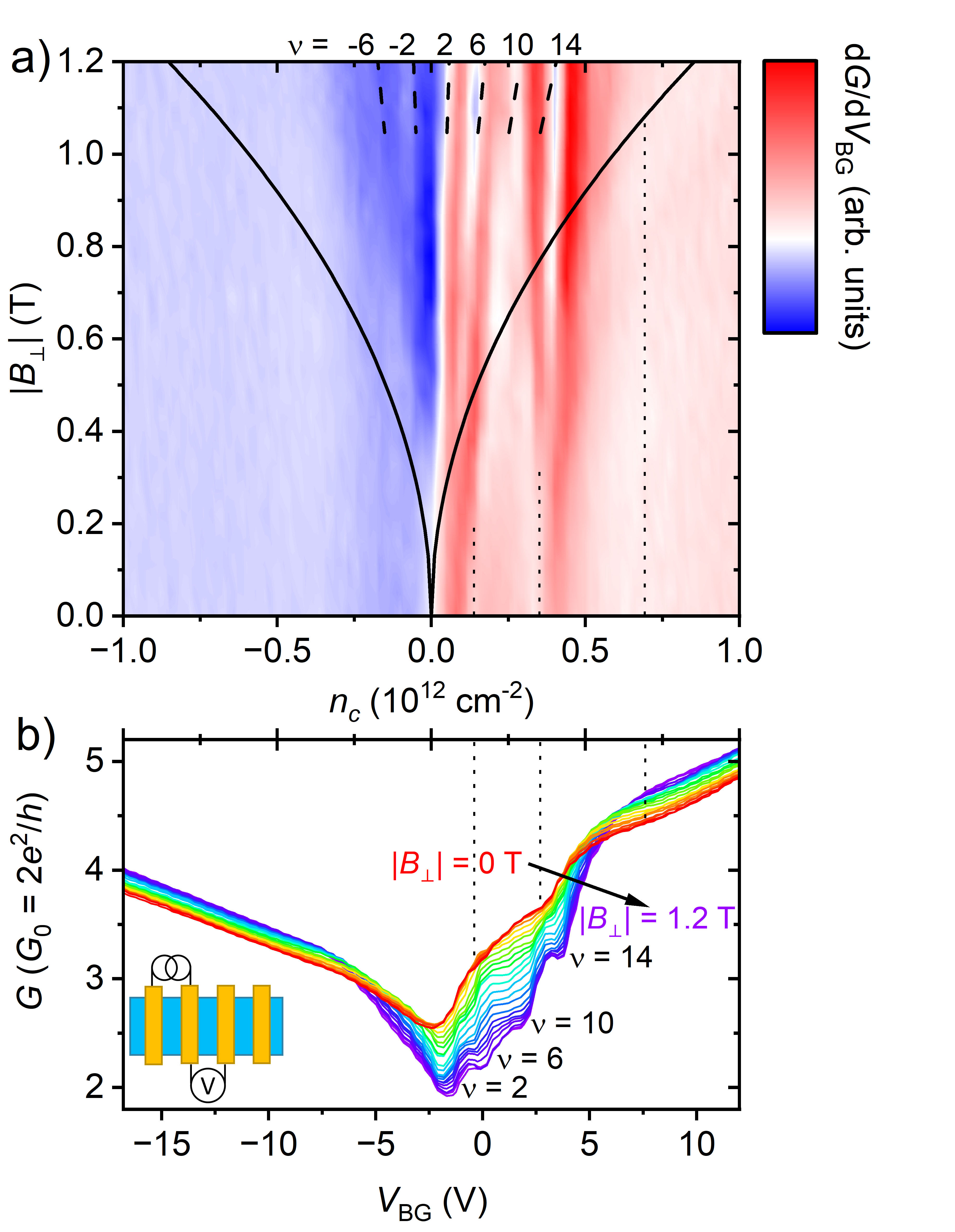}
    \caption{\small Landau fan diagram for D1, symmetrised with respect to $B_\perp$. \textbf{(a)} 2D map of transconductance for a single e-QPC as a function of contact carrier density, $n_{\text{c}}$, and perpendicular magnetic field strength, $|B_\perp|$. White regions imply conductance plateaus (d\textit{G}/d\textit{V}\textsubscript{BG} = 0), occurring due to confinement at low $B$ (dotted black lines), or quantum Hall transport at high $B$ (dashed black lines, labelled by filling factor $\nu$). Solid black line plotted with Equation \ref{eq:critB}. \textbf{(b)} Evolution of $G$ vs $n$, with increasing $|B_\perp|$. Dotted black lines indicate positions of low field plateaus and correspond to those seen in panel (a).}
    \label{fig:FanAndSpectro}
\end{figure}

In order to explore quantum Hall transport, we measure conductance through an individual e-QPC, while varying its carrier density, $n_{\text{c}}$, and steadily increasing perpendicular magnetic field strength, $B$. Plotting transconductance against $n_{\text{c}}$ and $B$ produces a Landau fan diagram, which is shown in Figure \ref{fig:FanAndSpectro}a (see further data in Supplementary Figure S6). At low field, faint white vertical regions are visible, corresponding to 1D subbands arising from confinement (dotted lines in Figure \ref{fig:FanAndSpectro}a,b); bias spectroscopy measurements were also performed on this e-QPC with the features only resolved at zero bias (Supplementary Figure S4). The transition from quantum confinement into the quantum Hall regime, where the 1D subbands of a ballistic conductor coalesce into well defined Landau Levels, occurs at a critical magnetic field strength, $B_\text{c}$, at which the cyclotron radius, $l_\text{c}$ is equal to half the constriction width i.e. $2l_{\text{c}} \approx W_{\text{c}}$ \cite{Datta-1995}. For a certain energy (represented by the filling factor, $\nu$) the field strength required for the transition is given by $W_{\text{c}} \approx 2l_{\text{c}} = 2l_{\text{B}}\sqrt{\nu/2}$ \cite{Terres-2016}, where $l_{\text{B}} = \sqrt{\hbar/eB}$ is the magnetic length \cite{Guimaraes-2012}. Combining the above gives the expression for the critical field,
\begin{equation}
    B_{\text{c}}=\frac{h}{eW_{\text{c}}}\sqrt{\frac{n}{\pi}},
    \label{eq:critB}
\end{equation}
with this transition indicated by the solid black line in Figure \ref{fig:FanAndSpectro}a. Above this transition the plateaus evolve into dispersive Landau levels at high magnetic field, each corresponding to a filling factor $\nu$. The white features observed for $B> 1.0$ T align closely with the expected Landau level positions, indicated by dashed black lines (Figure \ref{fig:FanAndSpectro}a) and consistent with the relation $B=nh/e\nu$ \cite{Zhang-2006, Peres-2006, Bolotin-2009}.

Both quantum Hall and low field data show lower conductance for hole doping (Figure \ref{fig:FanAndSpectro}), owing to the $p$-$n$ junction nature of the e-QPCs. Transmission probability for electron conductance is estimated from the average spacing between plateaus, in both regimes (Figure \ref{fig:FanAndSpectro}b). At low field (red trace), the average spacing gives a value of $T(e) = 0.28\pm0.04$, while at high field (purple trace) the extracted value is $T(e) = 0.26\pm0.03$. Finite $T(e)$ in the quantum Hall regime, where Landau levels are expected to be spaced by $4e^2/h$ for graphene, confirms that reduced plateau spacing arises due to scattering from disordered edges at the nanowire-graphene interface and the surrounding potential profile, which reduces the transmission probability of charge carriers injected via the e-QPCs. 

\section{Conclusions} \label{sec:Conc}

We demonstrate quantised conductance and quantum Hall transport at 1D magnetic nanowire-graphene interfaces that form e-QPCs, capable of spin injection and detection, in a high mobility graphene spin transistor. Consistent values of effective constriction widths are extracted using two distinct methods: (i) using the Landauer equation to analyse $G$ vs $k_{\text{F}}$ data with a transmission estimated from conductance steps, and (ii) estimating the energy spacing, $\Delta E$, between consecutive subbands via finite bias analysis. Across all data and devices, the transmission probability falls within the range $T = 0.08-0.30$. Explicit evidence of spin transport, detected in a representative device, confirms that spin injection occurs ballistically, resulting in spin signals that propagate over $> 10\,\mu$m distances. The presence of 1D subbands in spin injecting contacts is confirmed; weak features in the spin signal suggest that $\Delta R_{\text{NL}}$ depends on the position of the Fermi energy in relation to the subbands, but more evidence is needed. Applying a perpendicular magnetic field prompts a transition into the quantum Hall regime, which causes the emergence of Landau level plateaus at the expected filling factors of 4-fold degenerate graphene. These phenomena take place in a graphene spintronic device without the need to engineer a physical constriction within the graphene channel. The demonstration of ballistic spin injection via magnetic e-QPCs presents an encouraging step towards the development of low-power ballistic spintronics. 

\section{Methods}

\subsection{Device fabrication}
Devices investigated in this work comprise Van der Waals heterostructures of single layer graphene flakes, fully encapsulated by thin ($10-30$ nm) hBN flakes. Individual flakes are isolated via mechanical exfoliation, then stacked via the dry peel transfer technique \cite{Mayorov-2011}. A polymer hard mask is patterned over the heterostructure using electron beam lithography (EBL), which, followed by reactive ion etching, defines the transport channels (Figure \ref{fig:Schematic}a and Supplementary Figure S1a). Due to the selectivity of the etch recipe, the hBN and graphene flakes have different etch rates, leading to a small ledge of exposed graphene ($\leq$10 nm) forming at the edges of the channel; this is the point where the 1D contact is made (Supplementary Figure S1b) \cite{Guarochico-Moreira-2022}. A second EBL stage is performed to pattern nanowire electrodes into a polymer coating.  Ferromagnetic metal is deposited onto the sample, which adheres to the exposed pattern and forms the electrodes. For the devices in this work, the nanowires are 60 nm cobalt (Co) with a 20 nm Gold (Au) capping layer. All contacts have widths in the range $W_{\text{c}} = 150 - 350$ nm. Devices are stacked on 290 nm SiO$_2$/Si substrates, with the oxide layer acting as a dielectric between the highly doped silicon back gate and the flakes of 2D material on its surface. 

\subsection{Electrical characterisation}

Electrical measurements are performed in a high vacuum cryostat (pressures $<10^{-6}$ mbar), at temperatures between room temperature (300 K) and low temperature (20 K); temperature is lowered by a constant flow of liquid helium into the cryostat, then controlled by a thermocouple placed near the device. Devices are mounted between the poles of a rotating electromagnet, allowing us to apply both in-plane and out-of-plane magnetic fields, with a strength of up to 1.2 T. All transport measurements are performed with low frequency ($<20$ Hz) AC lock-in techniques, with the exception of bias spectroscopy (see below methods).

Carrier density in the graphene channel, $n$, is controlled by the application of a DC voltage, $V_{\text{BG}}$, to the p-doped silicon back gate, which changes the electric field between the gate and channel. In order to estimate $n$ in the graphene, the channel and Si back gate are treated as plates of a parallel capacitor, separated by two dielectrics: (i) the SiO$_2$ layer, with thickness $d_{\text{Si}} = 290$ nm, and (ii) bottom hBN flake of thickness $d_{\text{h}}$, determined via AFM (Supplementary Figure S1b). We apply the equation $n=C_{\text{g}} (V_{\text{BG}}-V_{\text{D}})/e$, where $V_{\text{D}}$ is the gate voltage at which the Dirac point occurs, and the capacitance is calculated as $C_{\text{g}} = \epsilon_0 \epsilon_{\text{h}} \epsilon_{\text{Si}}/(\epsilon_{\text{h}} d_{\text{Si}}+\epsilon_{\text{Si}} d_{\text{h}})$, where $\epsilon_0 = 8.85 \times 10^{-12}, \epsilon_{\text{h}}\sim3.8,$ and $\epsilon_{\text{Si}} = 3.9$ are the permittivities of free space, hBN (value used here is for bulk hBN \cite{Laturia-2018}), and SiO$_2$, respectively. In this interpretation, applying a positive (negative) voltage to the gate will induce negative (positive) charges in the graphene channel. Hence, by gating, we can access both the electron and hole transport regimes. 

Charge transport in the graphene channel is probed by measuring resistivity, $\rho$, in the standard 4-terminal configuration (Supplementary Figure S2a), while controlling channel carrier density, $n$, by the means explained above. The result is a Dirac curve (Supplementary Figure S2a). 

Transport at the 1D contact is measured via the 3-terminal configuration (Figure \ref{fig:Ballistic}c and Supplementary Figure S2d). $V_{\text{BG}}$ is again used to control the carrier density, but in this case a specific contact carrier density, $n_{\text{c}}$, is defined by the voltage at which minimum contact conductance occurs. All contacts give resistances in the range $R_{\text{c}} = 1 - 10$ k$\Omega$.

The charge diffusion coefficient, $D$, is calculated via the Einstein relation, $D = \sigma/e N$, with $N$ being the density of states, calculated as,

\begin{equation}
    N = \frac{\sqrt{g_s g_v n}}{\sqrt{\pi} \hbar v_{\text{F}}},
    \label{eq:DOS}
\end{equation}

\noindent
where $g_s$ and $g_v$ are the spin and valley degeneracy factors, respectively, and $v_{\text{F}}$ is the Fermi velocity. 

\subsection{Bias spectroscopy}

In the case of bias spectroscopy, a DC voltage source is used in series with an AC lock-in amplifier, such that we can measure the differential conductance of an e-QPC, $G = dI/dV$, while simultaneously applying a bias voltage between contact and channel, $V_{\text{Bias}}$. The amplitude of the AC signal used is chosen to be small, such that the magnitude of oscillations in $V_{\text{Bias}}$ is far below the energy scale of the features being resolved. Using an independent DC source, a voltage is applied to the silicon back gate, $V_{\text{BG}}$, which controls the carrier density in the channel, $n$. Figures \ref{fig:Ballistic}a, \ref{fig:Ballistic}b, and Supplementary Figures S3a, S4 show the results of these measurements, where each black line represents a measurement of $G$ as a function of $V_{\text{Bias}}$, at a fixed $V_{\text{BG}}$.

\section{Acknowledgments}
UK participants in Horizon Europe Project “2D Heterostructure Non-volatile Spin Memory Technology” (2DSPIN-TECH) are supported by UKRI grant number [10101734] (The University of Manchester). D.B. acknowledges the Engineering and Physical Sciences Research Council (EPSRC), Doctoral Training Partnership (DTP) for funding the PhD project (U.K). J.C.T.F. acknowledges support from the Consejo Nacional de Ciencia y Tecnología (México). V.H.G.M. acknowledges support from the Secretaría Nacional de Educación Superior, Ciencia y Tecnología (SENESCYT), for the Ph.D. scholarship under the program “Universidades de Excelencia 2014” (Ecuador). 

\section{Data availability}
Research data are available from the authors upon request.

\section{Competing interests}
The authors declare no competing interests.

\section{Author Contributions}
D.B. performed measurements, conducted data analysis, created the figures, and wrote the manuscript. J.C.T.F. performed measurements, contributed to data analysis, and contributed to the writing of the manuscript. V.H.G.M. Fabricated devices, performed measurements, and contributed to the writing of the manuscript. K.O. Contributed towards measurements. I.V.G. and T.T. contributed to the writing of the manuscript. I.V.J.M. conceptualised the work, contributed to data analysis, and contributed to the writing of the manuscript.

%\bibliography{bib}% Produces the bibliography via BibTeX.


\end{document}